\documentclass{ci}
\usepackage[T1]{fontenc}
\usepackage{shorthand}

\begin{document}
\bibliographystyle{agsm}

\title{Crowd \& Prejudice: An Impossibility Theorem for Crowd Labelling without a Gold Standard}

\numberofauthors{2} 
\author{
	\alignauthor Nicol\'{a}s Della Penna\\
      \affaddr{ANU NICTA}\\
       \affaddr{Locked bag 8001, 2601}\\
       \affaddr{Canberra, ACT, Australia}\\
       \email{nicolas.della-penna@anu.edu.au}
\alignauthor Mark D. Reid\\
       \affaddr{ANU NICTA}\\
       \affaddr{Locked bag 8001, 2601}\\
       \affaddr{Canberra, ACT, Australia}\\
       \email{mark.reid@anu.edu.au}
}

\maketitle
\begin{abstract}
A common use of crowd sourcing is to obtain labels for a dataset. Several algorithms have been proposed to identify uninformative members of the crowd so that their labels can be disregarded and the cost of paying them avoided. One common motivation of these algorithms is to try and do without any initial set of trusted labeled data. We analyse this class of algorithms as mechanisms in a game-theoretic setting to understand the incentives they create for workers. We find an impossibility result that without any ground truth,
and when workers have access to commonly shared 'prejudices' upon which they agree but are not informative of true labels, there is always equilibria where all agents report the prejudice. A small amount amount of gold standard data is found to be sufficient to rule out these equilibria.
\end{abstract}

\section{Introduction}

\begin{quote}
\emph{For ``the crowd'' is untruth.}-- Kierkegaard
\end{quote}

Precedent literature has proposed a large number of algorithms that take a set of data points labeled by a group of agents,
and try to estimate both the reliability of agents. These algorithms can be divided into two sets: those that leverage a small amount of gold standard (ground truth) data \cite{snow2008cheap,wauthierbayesian}, and those that do not \cite{dekel2009vox,raykar:2009,raykar2010learning,kumar2011modeling,Dekel:2009a,Yan:2010}. These algorithms that attempt to do without the need for gold standard, do so by using agreement among different labellers as indicative of correctness of a label. This agreement is either at the level of how to label of a given datapoint, as in most cases; or in how features map to labels, as in \cite{dekel2009vox}. To achieve this they place their trust on agents who provide labels that are consistent with the labels provided by other agents, or in the case where the same datapoint is not labeled twice, where the proposed feature to label mapping is consistent with other agents'  mapping of features to labels. It is often the case that labellers want to be seen as informed by those who are collecting the labels, as the labelling tasks soften pay and it is natural for those collecting the data to avoid the unnecessary cost of paying for labels from uninformed labellers.

We analyse the class of algorithms that do not use gold standard data as mechanisms in a game-theoretic setting, in order to understand the incentives they create for the agent providing the labels. We first present an impossibility result:
that without gold label data, and when workers have access to commonly shared 'prejudices' upon which they agree but which
are not informative of true labels, then there is always equilibria where the mechanism does not obtain the true labels from the informed workers, but rather all workers report the prejudice. We then consider how a small amount of gold data is generally
sufficient to render situations where the prejudice is reported by informed players as outside the equilibrium set.

One possible criticism of our work is that there is little interest in pointing out that when the assumptions of a statistical model (in this case, that agreement among labellers indicates correctness) do not hold the conclusions drawn from such a model can be misleading. Our argument, however, is more subtle than this: the incentives created by the natural applications of the model in its
 intended task undermine the very assumptions of the model, by creating incentives for players to agree on the labelling with others,
  irrespective of whether  they believe these  to be the true labels. 

To make the situation we have in mind more concrete and to clarify
how it defers from standard information cascades studied in economics, consider the hypothetical example of
a professor who assigns their teaching assistants to grade exams, without grading any themselves. The TAs may or may not know the topic at hand (be informed or uninformed) and they must provide a grade (label) to each exam they are assigned. If the TAs each grade a question on each exam and do so sequentially, so that for each previous answer in a given exam they can observe the grades other TAs have assigned to them, what in economics is referred to as an information cascade can occur. TAs grading later questions can look at the grade a student received for initial exam questions, and guess that the question they where assigned will receive a similar grade, instead of having to understand the answer to the question the student gave and how it relates to the correct 
 answer. 
 In contrast, we study a related but different situation, analogous to one, in which each TA grades (possibly overlapping) full exams, and they do so simultaneously without access to what the others are assigning. Note that if the TAs expect to be rewarded for agreement with others and if they believe others may use some prejudice to grade the exam, such as assigning higher grades to, say, students with neat hand writing or who use longer words, they might be motivated to also used said prejudice. 

An equivalent example can be considered in a crowd sourcing context. Suppose we ask for translations of a given word in language A to speakers of language B; a common prejudice for speakers of language B would be that if a word in language A sounds like the word in language B it must translate to that word. Even if bilingual speakers of A and B are present in the worker pool, if they believe this, the consensus label will be the similar sounding (but possibly incorrect) translation they may choose to report this prejudice as to continue to be employed in the translation task.

\section{Related Literature}

The study of learning algorithms from a mechanisms design perspective was initiated by \cite{dekel2008incentive} on a task they term ``incentive compatible regression learning'',  where agents care about the function that is learned and can report their observations to strategically manipulate it. In contrast to that model, our agents are not motivated to manipulate the learned function mapping examples to labels but are instead are motivated to be seen by the mechanism as capable and thus to continue to be employed as a source of labels for the task. In \cite{meir2010limits} a strategy-proof mechanisms where agents report labels and their objective is to maximize the accuracy of the learned classifier only on their subset of the data. 

Mechanisms designed to elicit subjective probabilities truthfully exploit richer action sets, where the action is not just reporting a label,
 but also reporting the distribution of labels the population will report. Examples of this type of mechanism
  is the Bayesian Truth Serum introduced in \cite{prelec2004bayesian}, and the extension of the Peer Prediction Method \cite{Miller2005} proposed by \cite{Witkowski2011}.

Our 'prejudice' can be thought of as 'extrinsic random variables' which allow agents to coordinate their decisions, models for the equilibrium of these have been extensively studied in the economics literature. For a recent review of the literature see \cite{shell2008sunspot}, for experimental evidence for the laboratory see \cite{duffy2005sunspots}

A rich literature on herding behaviour exists in economics, and is closely related to the model we examined but in a sequential instead of simultaneous setting, thus the externality that encourages the herding is of an informational nature instead of in our case where it directly affects payoffs. When agents arrive exogenously ordered sequence and can observe previous agents choices and can follow these, they can either avoiding paying the cost of acquiring information about the payoff of actions themselves or disregarding private information which they may posses, the classic papers in this stream are  \cite{banerjee1992simple,bikhchandani1992theory}. Experimental laboratory studies have been carried out looking at herding and information cascades, both in the laboratory \cite{cipriani2005herd} and in the internet \cite{drehmann2005herding} . 

For a recent multidisciplinary review of herding in humans from a cognitive neuroscience perspective see \cite{raafat2009herding}

\section{Setting}

Let $\Ycal=\{1,\ldots,K\}$ a set of labels. We consider a game between
the \emph{world,} a \emph{mechanism }$M$, and a set of \emph{agents}
$\Acal$. Each agent $a\in\Acal$ falls is of one of two types:
\emph{informed} or \emph{uninformed}. We denote the set of informed
agents by $\Acal_{I}\subseteq\Acal$. The goal of the mechanism is
to identify which agents that are informed. The goal of the agents is
to be identified as informed by the mechanism, even if they are not.

Each game is determined by a distribution $P$ over $\Ycal^{3}$ with
the random variables $(Y,U,I)\sim P$. Letting $I(\cdot;\cdot)$ denote
mutual information, we require that $P$ satisfy three conditions:

\begin{equation}
I(Y;U)=0
\end{equation}

\begin{equation}
I(Y;I)>0
\end{equation}

\begin{equation}
P(Y)\ne P(U)
\end{equation}
 
  We note that
conditions 1 and 2 are equivalent to requiring $P(Y,U,I)=P(U)P(Y)P(I|Y)$
and $P(Y,I)\ne P(Y)P(I)$, respectively. Intuitively, $Y$ is to be
interpreted as the {}``true'' label the mechanism is trying to
learn, $U$ is some uninformative signal about $Y$ that has a different
distribution to $Y$, and $I$ is an informative signal about $Y$.
It is assumed that $P(Y)$ is common knowledge to the mechanism and
all the agents.

The game is played by the world first secretly drawing $y\sim P(Y)$.
Every agent $a\in\Acal$ then receives an i.i.d. draw $u_{a}$ from
$P(U|Y=y)=P(U)$. In addition, informed agents receive a draw $i_{a}$
from $P(I|Y=y)$. The agents then each decide to report some $y_{a}\in\Ycal$
to the mechanism and from this the mechanism must try to determine
which agents are informed and which are not. 

Strategically, uninformed agents have two choices: they can play a
\emph{prejudiced} strategy and report $y_{a}=u_{a}$, or they can
\emph{randomise} and draw a new $y_{a}$ from $P(Y)$. Informed agents
can also play prejudice or randomise but, in addition, can also play
a \emph{truthful} strategy and report $y_{a}=i_{a}$. The decision
for an agent to be truthful, prejudiced, or random depends on the
mechanism. An informed agent may strategically decide to not be truthful
in order to maximise its chances of being identified as informed.

For our purposes, a mechanism is a function from a set of reports
$R=\{y_{a}:a\in\Acal\}$ to set of agents $\Acal_{M}\subseteq\Acal$
that the mechanism identifies as informed and truthful. That is, $M:\Ycal^{\Acal}\to2^{\Acal}$.
The goal of $M$ is to maximise the probability that $\Acal_{M}$
coincides with $\Acal_{I,T}$ the set of informed agents. It suffices
for $M$ to ensure that

\begin{equation}
p_{I,I}=P(a\in\Acal_{M}|a\in\Acal_{I,T})>\frac{1}{2}
\end{equation}

\begin{equation}
p_{U,U}=P(a\notin\Acal_{M}|a\notin\Acal_{I,T})>\frac{1}{2}
\end{equation}

since repeated independent samples will guarantee that $\left(\frac{p_{I,I}}{1-p_{I,I}}\right)^{n}\to\infty$
and $\left(\frac{p_{U,U}}{1-p_{U,U}}\right)^{n}\to0$ as the number
of labeled examples $n$ goes to infinity.

\section{Results}

We consider the class of mechanisms satisfying the above which succeed when a majority of players are informed and their reports truthful and uninformed players randomise. All proposed algorithms in the literature, to our knowledge, satisfy this elementary criterion.

\subsection{Equilibria}

Three Bayes-Nash equilibria of the game induced by these mechanisms are:

\begin{description}
\item[All randomise.] When all other agents are randomizing,
an agent is indifferent among all labels and thus also about randomising over them.  This is not a particularly robust equilibrium as deviation of two agents to prejudice is sufficient to cause all others players to deviate it, and a deviation of two informed agents to truthfulness is also sufficient to cause all other informed agents to improve their payoff by deviating to truthfulness. 
\item[Informed are truthful and uninformed randomise.] In this equilibrium the mechanism by and large works in the sense that the set of players that acts consistently in the same manner as the set of informed and truthful agents.
\item[Both play prejudice.] When all other plays play prejudice an agent's probability of having their labels found to coincide with others is maximized by playing prejudice, rewardless of whether they are informed or uninformed.

An interesting open question is how to select among these equilibria; while the fragility of the equilibrium where all agents randomize irrespective of type makes it an unlikely candidate it is unclear how to select among the other two. We conjecture that given equal sized populations of informed and uninformed agents the equilibrium selected will depend on the relative entropy of the prejudice and informed distribution, with the lower entropy distribution being more likely to be selected.

\end{description}

\subsection{Impossibility}

Any mechanism in the class defined above must fail in some equilibrium and for some distribution. Consider a situation where all agents are informed and play truthfully:  the mechanisms succeeds if and only if it identifies all agents as informed and truthful. Now consider a new situation, one in which the equilibrium where all agents play the prejudice occurs, and the distribution of the prejudice in this situation is identical to the distribution of the truth in the previous situation. Since the play observed in both games is identical from the perspective of the mechanism, it must designate all agents as informed and truthful in the second situation and fail, or it must have identified some agents as prejudiced in the first situation and fail. 

\subsection{Using a gold standard}

A mechanism that has access to sufficient gold standard data (a sample from the informed distribution) is enough for the impossibility result to no longer apply. In the situation where agents play the prejudice with high probability, the labels reported by those who are playing prejudice will contradict the labels the mechanism has access to and thus can be identified. The mechanism
 needs access to enough labels from the informative distribution to identify either an agent that is playing prejudice or one
 that is playing truthfully. It can then extrapolate to agents who have labeled points which it did not originally have labels for based on their agreement or disagreement with those agents it previously identified. This new set of players that has then been identified can be used to extrapolate the the players who labelled points in common with them, and  this procedure can be repeated until all players have been identified (this requires that there is sufficient overlap between the data points labeled by agents). The same logic can be adapted to algorithms that do not have players label points in common such as \cite{dekel2009vox} but the overlap then applies to  how features map to labels instead of how labels map to points. 
 
Interestingly, access to the prejudice is also sufficient to for the mechanism to succeed in that situation, as this can also be used to identify those who are playing prejudice with high probability when they overlap with the points to which the mechanism
 has access to the prejudice. The same overlapping procedure can then be used to reveal the strategies of the other players.

Since randomising provides the highest entropy to the sequence of play and thus is the hardest to distinguish from the true labels that a uninformed player can generate  the mechanism can identify players who play the prejudice in this situation faster than those that are randomizing.  Thus, there is no equilibrium where the mechanism uses the gold data where agents play prejudice. The gold data also guarantees that informed agents have a dominant strategy to be truthful. This implies the only equilibrium has informed agents playing truthfully and uninformed agents randomizing.

\section{Conclusion}

We consider algorithms that attempt to distinguish between informed and uninformed workers without using gold standard data, and show that when these algorithms are analyzed as mechanisms they can lead to equilibria where no agents truthfully reveal their private information about the label if they have access to it but rather report labels that are uninformative of the true label, but on which they can coordinate with other agents.  
In future research experimental work to identify wether the equilibria identified in the theoretical model occur, and to test theories of equilibrium selection if they do, would be extremely interesting. 

\section{Acknowledgments}
We would like to gratefully acknowledge Maureen Evans for help with editing the paper, and an anonymous reviewer for their helpful detailed comments. The idea that algorithms proposed for learning without gold standard data can be seen as models for herding behaviour come from a blog post by Paul Mineiro.  \cite{mineiro2011herd}

\section{Citations and References}

\bibliography{collective-intelligence}  

\end{document}